\documentclass[12pt,thmsa]{article}
\usepackage{sw20lart}

\input{tcilatex}
\begin{document}

\title{The Possibility of Factorizable Contextual Hidden Variable Theories}
\author{A. Shafiee\thanks{%
e-mail: sepafs@netware2.ipm.ac.ir}\qquad M. Golshani\thanks{%
e-mail: golshani@ihcs.ac.ir} \\
Institute for Theoretical Physics and Mathematics\\
(P. O. Box 19395-1795, Tehran, Iran)}
\maketitle

\begin{abstract}
Considering an extended type of Bohm's version of EPR thought experiment, we
derive Bell's inequality for the case of factorizable contextual hidden
variable theories which are consistent with the predictions of quantum
theory. Usually factorizability is associated with non-contextuality. Here,
we show that factorizability is consistent with contextuality, even for the
ordinary Bohm's version of the EPR thought experiment.
\end{abstract}

\section{Introduction}

In the derivation of almost all Bell inequalities the assumption is made
that all dynamical variables involved are not compatible. It is also a
common understanding that in thought experiments like Bohm's version [1] of
EPR's [2] (hereafter called EPRB), the violation of all Bell inequalities
are due to non-local effects between the two correlated particles. In the
case of systems involving two correlated particles, it is generally believed
that Bell's locality condition (i.e. factorizability) is equivalent to
non-contextuality and that non-contextual hidden variable theories are
incompatible with the standard quantum mechanics [3,4]. This kind of
incompatibility has been primarily discussed by Gleason [5], Bell [6] and
Kochen-Specker [7]. Nevertheless, the meaning of contextuality has not been
expanded very well. In the present paper, we extend the notion of
contextuality as originating from one of the following sources or both of
them :

1) State preparation of the system [8,9] ;

2) Measured value of a dynamical variable is dependent on what other
commuting dynamical variables are being measured along with it [3].

The significance of this distinction is made clear in section 2, where we
make use of a thought experiment for four compatible observables, resembling
the EPRB thought experiment, in which non-contextuality ( in the sense of
factorizability) is not violated. Considering the first version of
contextuality, we show that the possibility of having compatibility between
factorizable contextual hidden variable theories and quantum mechanics
exists. Then, in section 3 we argue that even for the EPRB thought
experiment this possibility remains open.

\section{An EPRB-type Thought experiment}

We consider a source of spin zero systems which emits two spin-1/2
particles. On its way, each particle meets two Stern-Gerlach apparatus (S.G$%
_{\text{1}}$ and S.G$_{\text{2}}$) at times $t_{1}$and $t_{2}$ ( $t_{2}>$ $%
t_{1}$), respectively. The spin of particle 1 (2) is being measured at
either $t_{1}$or $t_{2}$ along $\widehat{a}\ (\widehat{b})$ or $\widehat{%
a^{\prime }}\ (\widehat{b^{\prime }})$ respectively. We represent the
results of each measurement in the following way:

\[
\sigma _{a,1}^{(t_{1})}\rightarrow A_{1};\ \sigma _{a^{\prime
},1}^{(t_{2})}\rightarrow A_{2};\ \sigma _{b,2}^{(t_{1})}\rightarrow B_{1};\
\sigma _{b^{\prime },2}^{(t_{2})}\rightarrow B_{2} 
\]
where $\sigma _{a,1}^{(t_{1})}$ refers to the spin component of particle 1
measured along $\widehat{a}\ $at time $t_{1}$, and similarly for others.
Using the units of $\hbar /2,$ the quantities $A_{1,}B_{1},A_{2},$ and $%
B_{2} $ take the values $\pm 1.$

\FRAME{dtbpF}{5.1742in}{1.9761in}{0pt}{}{}{fig2.bmp}{\special{language
"Scientific Word";type "GRAPHIC";display "USEDEF";valid_file "F";width
5.1742in;height 1.9761in;depth 0pt;original-width 492.3125pt;original-height
176.1875pt;cropleft "0";croptop "1";cropright "1";cropbottom "0";filename
'D:/fig2.bmp';file-properties "XNPEU";}}

\bigskip

In a non-contextual hidden variable theory, we define non-contextuality as a
conjunction of the following two assumptions for four compatible dynamical
variables $\sigma _{a,1}^{(t_{1})},\ \sigma _{a^{\prime },1}^{(t_{2})},$ $%
\sigma _{b,2}^{(t_{1})}$and $\sigma _{b^{\prime },2}^{(t_{2})}$ :

\begin{quote}
\underline{$\mathbf{C}_{1}$}.\textbf{\ \ }The value obtained for the spin of
each particle at $t_{2}$ is independent of the \textit{prepararion} made for
the spin state of the same particle at an earlier time $t_{1}$(i.e. the
orientation of the Stern-Gerlach apparatus through which the particle passed
at $t_{1}$).
\end{quote}

Here, it has been taken for granted that the result obtained for a
particle's property at any time is independent of the presence of any
apparatus in the path of the same particle at a later time.

\begin{quote}
\underline{$\mathbf{C}_{2}$}. \ For the whole system, the result obtained
for each particle should be independent of the \textit{measurement} made on
the other particle at that time or at any other time. This is equivalent to
Bell's locality condition.
\end{quote}

We consider this experiment in the light of a hidden variable theory and
assume that the separation of the two particles 1 and 2 are space-like. Then
we make use of the following postulates:

\begin{quotation}
\underline{$\mathbf{P}_{1}$}. \ The spin state of the particles 1 and 2 is
described by a function of a collection of hidden variables called $\lambda $%
, which belongs to the space $\Lambda $. The parameter $\lambda $ contains
all the information which is necessary to specify the spin state of the
system. One can define a grand joint probability $p_{GJP}$ for the dynamical
variables $\sigma _{a,1}^{(t_{1})},\ \sigma _{a^{\prime },1}^{(t_{2})},$ $%
\sigma _{b,2}^{(t_{1})}$and $\sigma _{b^{\prime },2}^{(t_{2})}$ on the space 
$\Lambda ,$ representing the statistical informations which are possible to
obtain from the spin sate of the system.
\end{quotation}

Using $\mathbf{P}_{1},$ it would be possible to obtain the marginal
probabilities from $p_{GJP}$. For example, we have

\begin{equation}
p(A_{2},B_{2}|\widehat{a},\widehat{b},\widehat{a^{\prime }},\widehat{%
b^{\prime }},\lambda )=\stackunder{A_{1},B_{1}}{\dsum }A_{1}B_{1}%
\;p_{GJP}(A_{1},B_{1},A_{2},B_{2}|\widehat{a},\widehat{b},\widehat{a^{\prime
}},\widehat{b^{\prime }},\lambda )
\end{equation}

\begin{equation}
p(A_{1}|\widehat{a},\widehat{b},\lambda )=\stackunder{A_{2},B_{1},B_{2}}{%
\dsum }A_{2}B_{1}B_{2}\;p_{GJP}(A_{1},B_{1},A_{2},B_{2}|\widehat{a},\widehat{%
b},\widehat{a^{\prime }},\widehat{b^{\prime }},\lambda )
\end{equation}

The relation (1) gives the probability that the joint values of the spin
components of the particles 1 and 2 along $\widehat{a^{\prime }}\ $and $%
\widehat{b^{\prime }}$ at $t_{2}$ being $A_{2}$ and $B_{2}$ respectively,
assuming that $\lambda $ and the directions $\widehat{a},\widehat{b},%
\widehat{a^{\prime }},$and $\widehat{b^{\prime }}$ are completely specified.
Similarly, (2) gives the probability that the value of particle 1's spin
component along $\widehat{a}$ at $t_{1}$ being $A_{1},$ assuming that $%
\lambda $ and $\widehat{a}$ and $\widehat{b}$ are known. This probabilty is
independent of the directions $\widehat{a^{\prime }}\ $and $\widehat{%
b^{\prime }}$ related to time $t_{2}$.

\begin{quotation}
\underline{$\mathbf{P}_{2}$}. \ Any information about the values of the spin
components of particles 1 and 2, which are spatially separated, originates
from $\lambda $ (common causes) and is not related to any influence of one
particle over the other.
\end{quotation}

Due to $\mathbf{P}_{2},$ probability functions of the type (1) should be
factorizable:

\begin{equation}
p(A_{2},B_{2}|\widehat{a},\widehat{b},\widehat{a^{\prime }},\widehat{%
b^{\prime }},\lambda )=p(A_{2}|\widehat{a},\widehat{a^{\prime }},\lambda )\
p(B_{2}|\widehat{b},\widehat{b^{\prime }},\lambda )
\end{equation}
and for probability functions of the type (2) we have:

\begin{equation}
p(A_{1}|\widehat{a},\widehat{b},\lambda )=p(A_{1}|\widehat{a},\lambda )
\end{equation}

Our postulate $\mathbf{P}_{2}$ is equivalent to Bell's locality condition
and is also equivalent to $\mathbf{C}_{2}$ in a non-contextual theory.

Using $\mathbf{P}_{1}$ and $\mathbf{P}_{2}$, one can derive a Bell-type
inequality. This inequality is not violated by quantum mechanical
predictions. To obtain this inequality, we define

\begin{equation}
E_{1}^{(t_{1})}(\widehat{a},\lambda )=\stackunder{A_{1}}{\dsum }A_{1}\
p(A_{1}|\widehat{a},\lambda )
\end{equation}

\begin{equation}
E_{2}^{(t_{2})}(\widehat{b},\widehat{b^{\prime }},\lambda )=\stackunder{B_{2}%
}{\dsum }B_{2}\ p(B_{2}|\widehat{b},\widehat{b^{\prime }},\lambda )
\end{equation}

\begin{equation}
E_{12}(\widehat{a},\widehat{b},\widehat{b^{\prime }},\lambda
)=E_{1}^{(t_{1})}(\widehat{a},\lambda )\ E_{2}^{(t_{2})}(\widehat{b},%
\widehat{b^{\prime }},\lambda )
\end{equation}

Here, $E_{1}^{(t_{1})}(\widehat{a},\lambda )$ and $E_{2}^{(t_{2})}(\widehat{b%
},\widehat{b^{\prime }},\lambda )$ are, respectively, the average values of
the spin components of paticle 1 along $\widehat{a}$ at $t_{1}$and particle
2 along $\widehat{b^{\prime }}$ at $t_{2},$ and $E_{12}(\widehat{a},\widehat{%
b},\widehat{b^{\prime }},\lambda )$ represents the average value of the
product of the spin components of particles 1 and 2 along $\widehat{a}$ and $%
\widehat{b^{\prime }},$ respectively.

Now, one can show [10] that if the variables $x,\ y,\ x^{\prime },\
y^{\prime }$ are confined to the interval $\left[ -1,1\right] $, then there
exists a function $S$ defined by

\[
S=xy+xy^{\prime }+x^{\prime }y-x^{\prime }y^{\prime } 
\]
which lies in the interval $\left[ -2,2\right] \ $. Here, we take

\[
x=E_{1}^{(t_{1})}(\widehat{a},\lambda );\ y=E_{2}^{(t_{2})}(\widehat{b},%
\widehat{b^{\prime }},\lambda );\ x^{\prime }=E_{1}^{(t_{2})}(\widehat{a},%
\widehat{a^{\prime }},\lambda );\ y^{\prime }=E_{2}^{(t_{1})}(\widehat{b}%
,\lambda ) 
\]
where all of theses variables lie in the interval $\left[ -1,1\right] .$
Thus, we have

\begin{equation}
-2\leq E_{12}(\widehat{a},\widehat{b},\lambda )+E_{12}(\widehat{a},\widehat{b%
},\widehat{b^{\prime }},\lambda )+E_{12}(\widehat{a},\widehat{b},\widehat{%
a^{\prime }},\widehat{b^{\prime }},\lambda )-E_{12}(\widehat{a},\widehat{b},%
\widehat{a^{\prime }},\lambda )\leq 2
\end{equation}

Multiplying through the probability density $\rho (\lambda )$ and
integrating over $\Lambda $ ( $\int_{\Lambda }\rho (\lambda )d\lambda =1$ ),
we get the following inequality at the quantum level

\begin{equation}
-2\leq \langle \sigma _{a,1}^{(t_{1})}\sigma _{b,2}^{(t_{1})}\rangle
+\langle \sigma _{a,1}^{(t_{1})}\sigma _{b^{\prime },2}^{(t_{2})}\rangle
+\langle \sigma _{a^{\prime },1}^{(t_{2})}\sigma _{b^{\prime
},2}^{(t_{2})}\rangle -\langle \sigma _{a^{\prime },1}^{(t_{2})}\sigma
_{b,2}^{(t_{1})}\rangle \leq 2
\end{equation}
where, e.g., we have set the quantum expectation values as

\begin{equation}
\langle \sigma _{a,1}^{(t_{1})}\sigma _{b^{\prime },2}^{(t_{2})}\rangle
=\dint_{\Lambda }\ E_{12}(\widehat{a},\widehat{b},\widehat{b^{\prime }}%
,\lambda )\ \rho (\lambda )d\lambda
\end{equation}

From quantum mechanics we have (see appendix )

\[
\langle \sigma _{a,1}^{(t_{1})}\sigma _{b,2}^{(t_{1})}\rangle =-\cos \theta
_{ab};\langle \sigma _{a,1}^{(t_{1})}\sigma _{b^{\prime
},2}^{(t_{2})}\rangle =-\cos \theta _{ab}\cos \theta _{bb^{\prime }}; 
\]

\[
\ \langle \sigma _{a^{\prime },1}^{(t_{2})}\sigma _{b,2}^{(t_{1})}\rangle
=-\cos \theta _{ab}\cos \theta _{aa^{\prime }};\ \langle \sigma _{a^{\prime
},1}^{(t_{2})}\sigma _{b^{\prime },2}^{(t_{2})}\rangle =-\cos \theta
_{ab}\cos \theta _{aa^{\prime }}\cos \theta _{bb^{\prime }} 
\]
where $\theta _{kl}$ is the angle between $\widehat{k}$ and $\widehat{l}\ (%
\widehat{k},\widehat{l}=\widehat{a},\widehat{b},\widehat{a^{\prime }},%
\widehat{b^{\prime }})$. Inserting these relations into (9), one obtains

\[
-2\leq -\cos \theta _{ab}-\cos \theta _{ab}\cos \theta _{bb^{\prime }}-\cos
\theta _{ab}\cos \theta _{aa^{\prime }}\cos \theta _{bb^{\prime }}+\cos
\theta _{ab}\cos \theta _{aa^{\prime }}\leq 2 
\]

This is never violated for any choice of $\theta _{ab},\theta _{bb^{\prime
}},$ and $\theta _{aa^{\prime }}$. Thus, once\textbf{\ }$\mathbf{P}_{1}$ and 
$\mathbf{P}_{2}$ are assumed, there is no incompatibility between a hidden
variable theory and quantum mechanics as far as this version of EPRB thought
experiment is concerned.

Here, we have not made any use of $\mathbf{C}_{1},$but $\mathbf{C}_{2}$ is
applied in the form of $\mathbf{P}_{2}.$ This means that a particle is
influenced by its past, as can been seen from the relations (6) and (7).
Now, we try to see what happens when we impose $\mathbf{C}_{1}.$ We impose $%
\mathbf{C}_{1}$ in the form of the following postulate:

\begin{quotation}
\underline{$\mathbf{P}_{3}$}.\ The statistical results obtained for a spin
component of a particle is independent of the sort of \textit{preparation }%
made for the spin state of the same particle at an earlier time. \quad
\end{quotation}

Using this postulate, one can write the relations (6) and (7) in the
following forms

\begin{equation}
E_{2}^{(t_{2})}(\widehat{b},\widehat{b^{\prime }},\lambda )=E_{2}^{(t_{2})}(%
\widehat{b^{\prime }},\lambda )=\stackunder{B_{2}}{\dsum }B_{2}\ p(B_{2}|%
\widehat{b^{\prime }},\lambda )
\end{equation}

\begin{equation}
E_{12}(\widehat{a},\widehat{b},\widehat{b^{\prime }},\lambda
)=E_{1}^{(t_{1})}(\widehat{a},\lambda )\ E_{2}^{(t_{2})}(\widehat{b^{\prime }%
},\lambda )
\end{equation}

As a consequence of the conjunction of $\mathbf{P}_{1}$ and $\mathbf{P}_{3}$
one gets (1) in the form

\begin{equation}
p(A_{2},B_{2}|\widehat{a^{\prime }},\widehat{b^{\prime }},\lambda )=%
\stackunder{A_{1},B_{1}}{\dsum }A_{1}B_{1}\;p_{GJP}(A_{1},B_{1},A_{2},B_{2}|%
\widehat{a},\widehat{b},\widehat{a^{\prime }},\widehat{b^{\prime }},\lambda )
\end{equation}

Now, by combining $\mathbf{P}_{1,}$ $\mathbf{P}_{2}$ and $\mathbf{P}_{3}$,
the inequality (8) takes the form

\begin{equation}
-2\leq E_{12}(\widehat{a},\widehat{b},\lambda )+E_{12}(\widehat{a},\widehat{%
b^{\prime }},\lambda )+E_{12}(\widehat{a^{\prime }},\widehat{b^{\prime }}%
,\lambda )-E_{12}(\widehat{a^{\prime }},\widehat{b},\lambda )\leq 2
\end{equation}

In the integrated form of this inequality, the quantum expectations are
defined in the following form

\begin{equation}
\langle \sigma _{a,1}^{(t_{1})}\sigma _{b^{\prime },2}^{(t_{2})}\rangle
=\dint_{\Lambda }\ E_{12}(\widehat{a},\widehat{b^{\prime }},\lambda )\ \rho
(\lambda )d\lambda
\end{equation}

The quantum expectation values in (15) are found to be

\begin{equation}
\langle \sigma _{a,1}^{(t_{1})}\sigma _{b^{\prime },2}^{(t_{2})}\rangle
=-\cos \theta _{ab^{\prime }}
\end{equation}
and similarly for other expectation values. Insertion of these expectation
values in the integrated form of (14) leads to the violation of it at the
quantum level. Thus, the postulate $\mathbf{P}_{3}$ leads to inconsistency
with quantum mechanics. But, the possibility of consistency between a
contextual factorizable hidden variable theory with quantum mechanics
remains open. Here, the contextuality is only used as the negation of $%
\mathbf{C}_{1},$which means that the preparation of the spin state of each
particle cannot be excluded from our calculations.

\section{\-EPRB Thought Experiment}

The combination of $\mathbf{P}_{1}$ and $\mathbf{P}_{3}$ in the previous
experiment, provides the possibility of conversion of our thought experiment
to that of EPRB. We show this case by taking the $\stackunder{%
t_{2}\rightarrow t_{1}}{\lim }($ $\mathbf{P}_{1}+\mathbf{P}_{3})=\mathbf{P}%
_{4}$ which can be stated in the following terms:

\begin{quotation}
\underline{$\mathbf{P}_{4}$}. There exists a limit for $t_{2}\rightarrow
t_{1},$ under which the conjunction of the two postulates $\mathbf{P}_{1}$
and $\mathbf{P}_{3}$ remains valid.
\end{quotation}

This justifies relations like (13) for the EPRB case. Now, we have

$\qquad \qquad \qquad $%
\[
\stackunder{t_{2}\rightarrow t_{1}}{\lim }p(\sigma _{a^{\prime
},1}^{(t_{2})}=A_{2},\sigma _{b^{\prime },2}^{(t_{2})}=B_{2}|\widehat{%
a^{\prime }},\widehat{b^{\prime }},\lambda )= 
\]

\[
\stackunder{t_{2}\rightarrow t_{1}}{\lim }\stackunder{A_{1}B_{1}}{\dsum }%
A_{1}B_{1}p_{GJP}(\sigma _{a,1}^{(t_{1})}=A_{1},\sigma
_{b,1}^{(t_{1})}=B_{1},\sigma _{a^{\prime },1}^{(t_{2})}=A_{2},\sigma
_{b^{\prime },2}^{(t_{2})}=B_{2}|\widehat{a},\widehat{b},\widehat{a^{\prime }%
},\widehat{b^{\prime }},\lambda ) 
\]

\begin{equation}
=p(\sigma _{a^{\prime },1}^{(t_{1})}=A_{2},\sigma _{b^{\prime
},2}^{(t_{1})}=B_{2}|\widehat{a^{\prime }},\widehat{b^{\prime }},\lambda )\;
\end{equation}

Similarly, combining $\mathbf{P}_{4}$ and $\mathbf{P}_{2}$ leads to Bell's
inequality in the form of (14). Here, $\mathbf{C}_{1}$ makes no sense, but $%
\mathbf{C}_{2}$ is used in the form of $\mathbf{P}_{2}$. However, we use $%
\mathbf{C}_{1}^{\prime }$ in the following form :

\begin{quotation}
\underline{$\mathbf{C}_{1}^{\prime }$}. \ For a system consisting of a pair
of particles 1 and 2, the spin state of the system is independent of the
parameters which might be involved as a consequence of the \textit{state
preparation} of the system, and the value of a spin component along an
arbitrary direction is completely determined by $\lambda .$
\end{quotation}

This assumption has a more comprehensive character than $\mathbf{C}_{1}$,
because $\mathbf{C}_{1}$ can be concluded from $\mathbf{C}_{1}^{\prime },$
as a speciall case.

Now, having introduced non-contextuality in a broader sense, we logically
conclude that since the violation of $\mathbf{P}_{3}$ leads to the violation
of $\mathbf{P}_{4}$, the violation of the integrated form of (14) in the
EPRB experiment should be a result of the violation of $\mathbf{P}_{4}.$
But, what does this mean?

This means that the conjunction of $\mathbf{P}_{1}$ and $\mathbf{P}_{3}$ is
not valid under the limit of $t_{2}\rightarrow t_{1},$ so that the relation
(17) can not be correct.The conjunction of $\mathbf{P}_{1}$ and $\mathbf{P}%
_{3}$, which leads to relations like (13), indicates that the possible
values of a pair of spin component related to two particles are completely
determined by $\lambda $ , when the measuring directions of the
corresponding spin components are specified, and are independent of the
factors which can be introduced by the preparation of the spin state of each
particle at an earlier time. This attitude which led ultimately to conflict
with the quantum mechanical predictions at the statistical level, was caused
by $\mathbf{P}_{3}$. The postulate $\mathbf{P}_{4}$, which is defined by $%
\stackunder{t_{2}\rightarrow t_{1}}{\lim }($ $\mathbf{P}_{1}+$ $\mathbf{P}%
_{3})$, indicates that the foregoing argument remains valid, when the spin
state of each particle can be in principle prepared simultaneously at two
differnt directions. Thus, the violation of $\mathbf{P}_{3}$ ( in the
previous section) and of $\mathbf{P}_{4}$ ( in this section) both mean that
the statistical result of any measurement should be locally assessed only in
the context of the preparation factors, since there is no reason for the
violation of $\mathbf{P}_{2}$ in both cases.

As a consequence of the violation of $\mathbf{P}_{4},$ we can use
contextuality as the negation of $\mathbf{C}_{1}^{\prime },$ which means
that the spin state of each particle cannot be independent of the
preparation conditions. Thus, (17) is not valid and we have

\begin{eqnarray}
\ p(\sigma _{a^{\prime },1}^{(t_{1})} &=&A_{2}^{\prime },\sigma _{b^{\prime
},2}^{(t_{1})}=B_{2}^{\prime }|X_{1}(\widehat{a^{\prime }}),X_{2}(\widehat{%
b^{\prime }}),\lambda )  \nonumber \\
&\neq &p(\sigma _{a^{\prime },1}^{(t_{1})}=A_{2},\sigma _{b^{\prime
},2}^{(t_{1})}=B_{2}|X_{1}(\widehat{a},\widehat{a^{\prime }}),X_{2}(\widehat{%
b},\widehat{b^{\prime }}),\lambda )
\end{eqnarray}
where

$\qquad \qquad p(\sigma _{a^{\prime },1}^{(t_{1})}=A_{2},\sigma _{b^{\prime
},2}^{(t_{1})}=B_{2}|X_{1}(\widehat{a},\widehat{a^{\prime }}),X_{2}(\widehat{%
b},\widehat{b^{\prime }}),\lambda )=\qquad $

\begin{equation}
\stackunder{A_{1}B_{1}}{\dsum }A_{1}B_{1}p_{GJP}(\sigma
_{a,1}^{(t_{1})}=A_{1},\sigma _{b,1}^{(t_{1})}=B_{1},\sigma _{a^{\prime
},1}^{(t_{1})}=A_{2},\sigma _{b^{\prime },2}^{(t_{1})}=B_{2}|X_{1}(\widehat{a%
},\widehat{a^{\prime }}),X_{2}(\widehat{b},\widehat{b^{\prime }}),\lambda )
\end{equation}

Here, $X_{1}(\widehat{a},\widehat{a^{\prime }})\ (X_{2}(\widehat{b},\widehat{%
b^{\prime }}))$ defines the particle 1 (2) in the context of $\widehat{a}$
and $\widehat{a^{\prime }}$ ($\widehat{b}$ and $\widehat{b^{\prime }}),$%
which shows the state dependence of particle 1 (2) on the preparation
factors $\widehat{a}$ and $\widehat{a^{\prime }}$ ($\widehat{b}$ and $%
\widehat{b^{\prime }})$. Similarly, $X_{1}(\widehat{a^{\prime }})\ (X_{2}(%
\widehat{b^{\prime }}))$ defines the state of particle 1 (2) in the context
of $\widehat{a^{\prime }}$ $(\widehat{b^{\prime }})$, which includes the
preparation effects on the same particle. The above state dependences hold
locally.

The failure of (17) for the EPRB case has been sometimes interpreted as a
failure of the definition of $p_{GJP}$ for the case of incompatible
observables ( see for example refs. [11]-[13]). We think this point of view
is not essential, since in principle, the definition of $p_{GJP}$ at a
hidden variable level is tenable.

\[
\]
{\Large Appendix\hspace{-0.05cm}\ } 
\[
\]
In order to calculate the quantum mechanical expectation values in our
proposed experiment, it is essential to define an original joint probability
function which describes the possible outcomes of the spin components of the
particles 1 and 2 at $t_{1}$ and $t_{2}$ :

\[
P(A_{1},B_{1},A_{2},B_{2}|\widehat{a},\widehat{b},\widehat{a^{\prime }},%
\widehat{b^{\prime }},\psi _{0})=\mid \langle \psi
_{0}|u_{a,1}^{(t_{1})}(\pm ),u_{b,2}^{(t_{1})}(\pm )\rangle \mid ^{2} 
\]

\begin{equation}
\times \mid \langle u_{a,1}^{(t_{1})}(\pm ),u_{b,2}^{(t_{1})}(\pm
)|u_{a^{\prime },1}^{(t_{2})}(\pm ),u_{b^{\prime },2}^{(t_{2})}(\pm )\rangle
|^{2}  \tag*{(A.1)}
\end{equation}
where the outcomes $A_{1},B_{1},A_{2}$ and $B_{2}$ take the values $\pm 1$.
In (A.1), $\mid \psi _{0}\rangle $ is the singlet state of the source and is
defined as a linear combination of two base vectors, $\mid z+\rangle $ and $%
\mid z+\rangle $, which correspond to the two eigenstates of $\sigma _{z}$ :

\begin{equation}
\mid \psi _{0}\rangle =\left[ \mid z+\rangle _{1}\otimes \mid z-\rangle
_{2}-\mid z-\rangle _{1}\otimes \mid z+\rangle _{2}\right]  \tag*{(A.2)}
\end{equation}

The quantum states (spin states ) for two particles along an arbitrary
direction relative to the z-axis, at $t_{1}$ or $t_{2}$ is represented by 
\[
\mid u_{m,1}^{(t_{j})}(\pm ),u_{n,2}^{(t_{j})}(\pm )\rangle =\ \mid
u_{m,1}^{(t_{j})}(\pm )\rangle \otimes \mid u_{n,2}^{(t_{j})}(\pm )\rangle 
\]
where $j=1,2$ ; $m=a,a^{\prime }$ and $n=b,b^{\prime }$. The individual spin
states for particle 1 are defined as

\[
\mid u_{m,1}^{(t_{j})}+\rangle =\cos \frac{\widehat{m}}{2}\mid z+\rangle
_{1}+\sin \frac{\widehat{m}}{2}\mid z-\rangle _{1}, 
\]

\begin{equation}
\mid u_{m,1}^{(t_{j})}-\rangle =-\sin \frac{\widehat{m}}{2}\mid z+\rangle
_{1}+\cos \frac{\widehat{m}}{2}\mid z-\rangle _{1},  \tag*{(A.3)}
\end{equation}
and similarly for particle 2.

Now, the quantum mechanical counterpart of relation (10) is obtained in the
following way

\begin{equation}
\langle \sigma _{a,1}^{(t_{1})}\sigma _{b^{\prime },2}^{(t_{2})}\rangle =%
\stackunder{A_{1}B_{2}}{\dsum }A_{1}B_{2}\ P(A_{1},B_{2}|\widehat{a},%
\widehat{b},\widehat{b^{\prime }},\psi _{0})  \tag*{(A.4)}
\end{equation}
where

\begin{equation}
\ P(A_{1},B_{2}|\widehat{a},\widehat{b},\widehat{b^{\prime }},\psi _{0})=%
\stackunder{A_{2}B_{1}}{\dsum }\ P(A_{1},B_{1},A_{2},B_{2}|\widehat{a},%
\widehat{b},\widehat{a^{\prime }},\widehat{b^{\prime }},\psi _{0}) 
\tag*{(A.5)}
\end{equation}
is the marginal joint probability of the results $A_{1}$ and $B_{2}$. Using
the relations (A.1)-(A.3), the relation (A.5) yields

\begin{equation}
\ P(A_{1},B_{2}|\widehat{a},\widehat{b},\widehat{b^{\prime }},\psi _{0})=%
\frac{1}{4}\left[ 1-A_{1}B_{2}\cos \theta _{ab}\cos \theta _{bb^{\prime
}}\right]  \tag*{(A.6)}
\end{equation}

Inserting (A.6) into (A.4), one gets

\[
\langle \sigma _{a,1}^{(t_{1})}\sigma _{b^{\prime },2}^{(t_{2})}\rangle
=-\cos \theta _{ab}\cos \theta _{bb^{\prime }} 
\]

The other quantum expectation values are obtained in a similar way.

\end{document}